\documentstyle[12pt,epsfig]{article}

\def\bge{\begin{equation}}
\def\ene{\end{equation}}
\def\bg{\begin{eqnarray}}
\def\en{\end{eqnarray}}
\def\nn{\nonumber}

\begin{document}

\begin{center}
{\bf THE QUARK-MESON COUPLING MODEL\footnote{
Invited talk at the XIV International Seminar on High Energy Physics 
Problems ``Relativistic Nuclear Physics and Quantum Chromodynamics'', 
Dubna, 17-22 August, 1998
}}

\vskip 5mm
K. Saito$^{\dag}$

\vskip 5mm

{\small {\it
Physics Division, Tohoku College of Pharmacy, \\
Sendai 981-8558, Japan 
}}
\\
$\dag$ {\it
E-mail: ksaito@nucl.phys.tohoku.ac.jp 
}
\end{center}

\vskip 5mm

\begin{center}
\begin{minipage}{150mm}
\centerline{\bf Abstract}
We review the quark-meson coupling model, in which
the quark degrees of freedom are explicitly involved to describe
the properties of not only nuclear matter but also finite nuclei.  
Then, we present the electric and magnetic form factors
for the proton bound in specific orbits for several closed-shell
nuclei. 

\vskip 5mm

\end{minipage}
\end{center}

\vskip 10mm

\section{Introduction}

Whether or not quark degrees of freedom play a significant role 
in nuclei is one of the central questions in nuclear physics. 
Recently, tremendous efforts have been devoted to the study of medium 
modifications of hadron properties in a nucleus  
(which are in some sense precursors of the QCD phase 
transition)~\cite{qm97}.  
At present lattice simulations have mainly been  
performed for finite temperature ($T$), with zero chemical 
potential.  
Therefore, many authors have investigated hadron properties at finite 
nuclear densities ($\rho_B$) using effective 
theories~[1--6].  

In this paper, we will introduce one of these effective theories, 
namely the quark-meson coupling (QMC) model~\cite{guichon,qmc1,qmc2}, 
and then report the medium modification of the electromagnetic form 
factors of the bound proton in finite nuclei.   

\section{The quark-meson coupling model}
\subsection{Effect of nucleon structure}
Let us suppose that a free nucleon (at the origin) consists of three light 
(u and d) quarks under a (Lorentz scalar) confinement potential, $V_c$.  
Then, the Dirac equation for the quark field $\psi_q$ is given by 
\bge
[ i\gamma\cdot\partial - m_q - V_c(r) ] \psi_q(r) = 0 , 
\label{dirac1}
\ene
where $m_q$ is the bare quark mass.  

Next we consider how Eq.(\ref{dirac1}) is modified when the nucleon is bound 
in static, uniformly distributed (iso-symmetric) nuclear matter.  
In the QMC model~\cite{guichon,qmc1,qmc2} it is assumed that 
each quark feels scalar $V_s^q$ and vector $V_v^q$ potentials, which are 
generated by the
surrounding nucleons, as well as the confinement potential. 
Since the typical distance between two nucleons around normal nuclear 
matter density 
($\rho_0 = 0.15$ fm$^{-3}$) is surely larger than the typical size of the 
nucleon (the radius $R_N$ is $\sim$ 0.8 fm), the interaction (except for 
the short-range part) between the nucleons 
should be colour singlet, namely a meson-exchange potential.  Therefore, this 
assumption seems appropriate when the baryon density $\rho_B$ is not high.  
If we use the mean-field approximation (MFA) for the meson fields, 
Eq.(\ref{dirac1}) may be rewritten as 
\bge
[ i\gamma\cdot\partial - (m_q - V_s^q) - V_c(r) 
 - \gamma_0 V_v^q ] \psi_q(r) = 0 . 
\label{dirac2}
\ene
The potentials generated by the medium are constants because the matter 
distributes uniformly. As the nucleon is static, the time-derivative 
operator in the Dirac equation can be 
replaced by the quark energy $-i \epsilon_q$.  
By analogy with the procedure applied to the nucleon
in QHD~\cite{walecka}, if we introduce the 
effective quark mass by $m_q^{\star} = m_q - V_s^q$, the Dirac equation 
(\ref{dirac2}) can be rewritten in the same form as that in free space 
with the mass $m_q^{\star}$ and the energy $\epsilon_q - V_v^q$, instead of 
$m_q$ and $\epsilon_q$. 
In other words, the vector interaction has {\em no effect 
on the nucleon structure} except for an overall phase in the quark wave 
function, which gives a shift in the nucleon energy.  This fact 
{\em does not\/} depend on how to choose the confinement potential $V_c$.  
Then, the nucleon energy at rest in the medium is given by 
$E_N = M_N^{\star}(V_s^q) + 3V_v^q$, 
where the effective nucleon mass $M_N^{\star}$ depends on {\em only the 
scalar potential}.  

Now we extend this idea to finite nuclei. 
The solution of the general problem of a composite, quantum particle 
moving in background scalar and vector fields that vary with position is 
extremely difficult.  One has, however, a chance to solve the particular 
problem of interest to us, namely light quarks confined in a nucleon which is 
itself bound in a finite nucleus, only because the nucleon motion is 
relatively slow and the quarks highly relativistic. Thus the 
Born-Oppenheimer approximation is naturally suited to 
the problem. 
Our approach in Ref.\cite{qmc1} was to start with a classical nucleon 
and to allow its internal structure 
to adjust to minimise the energy of three quarks in the ground-state of 
a system under constant scalar and vector fields, 
with values equal to those at the centre of the nucleon. 
Having solved the problem using the meson fields at the centre of 
the nucleon, 
one can use perturbation theory to correct for the variation of the 
scalar and vector fields across the nucleon bag. In first order perturbation 
theory only the spatial components of the vector potential 
give a non-vanishing contribution. 
This extra term is a correction to the spin-orbit force~\cite{qmc1}. 

As shown in Refs.\cite{qmc1,qmc2}, the basic result in the QMC model 
is that, in the scalar and vector  
meson fields, the nucleon behaves essentially as a point-like 
particle with an effective mass 
$M_N^{\star}$, which depends on the position through only the scalar  
field, moving in a vector potential. 

Let us suppose that the scalar and vector potentials in Eq.(\ref{dirac2}) are 
mediated by the $\sigma$ and $\omega$ mesons, and introduce their 
mean-field values, which now depend on position ${\vec r}$, 
by $V_s^q({\vec r}) = g_{\sigma}^q \sigma({\vec r})$ and 
$V_v^q({\vec r}) = g_{\omega}^q \omega({\vec r})$, respectively, where 
$g_{\sigma}^q$ ($g_{\omega}^q$) is the coupling constant of the quark-$\sigma$ 
($\omega$) meson.  Furthermore, we shall add the isovector vector meson  
$\rho$ and the Coulomb field $A({\vec r})$ to describe finite nuclei 
realistically~\cite{qmc1,qmc2}.  
Then, the effective Lagrangian density for finite nuclei, involving the quark 
degrees of freedom in the nucleon and the (structureless) meson fields  
in MFA, would be given by~\cite{qmc1,qmc2}
\bg
{\cal L}_{QMC-I}&=& \overline{\psi} [i \gamma \cdot \partial 
- M_N^{\star} - g_\omega \omega \gamma_0 
- g_\rho \frac{\tau^N_3}{2} b \gamma_0 
- \frac{e}{2} (1+\tau^N_3) A \gamma_0 ] \psi \nn \\
&-& \frac{1}{2}[ (\nabla \sigma)^2 + 
m_{\sigma}^2 \sigma^2 ] 
+ \frac{1}{2}[ (\nabla \omega)^2 + m_{\omega}^2 \omega^2 ] 
+ \frac{1}{2}[ (\nabla b)^2 + m_{\rho}^2 b^2 ] 
+ \frac{1}{2} (\nabla A)^2 , 
\label{qmclag}
\label{qmc-1}
\en
where $\psi$ and $b$ are respectively the nucleon and the $\rho$ fields. 
$m_\sigma$, $m_\omega$ and $m_{\rho}$ are respectively 
the (constant) masses of the $\sigma$, $\omega$ and $\rho$ mesons. 
$g_\omega$ and $g_{\rho}$ are respectively the $\omega$-N and $\rho$-N 
coupling constants, which are given by 
$g_\omega = 3 g_\omega^q$ and $g_\rho = g_\rho^q$ (where $g_\rho^q$ is 
the quark-$\rho$ coupling constant).  
We call this model the QMC-I model~\cite{qmc2}.  
If we define the field-dependent $\sigma$-N coupling 
constant $g_\sigma(\sigma)$ by
\bge
M_N^{\star}(\sigma({\vec r})) \equiv M_N - g_\sigma(\sigma({\vec r})) 
\sigma({\vec r}) , \label{coup}
\ene
where $M_N$ is the free nucleon mass, it is easy to compare with 
QHD~\cite{walecka}.  
The difference between QMC-I  
and QHD lies only in the coupling constant $g_\sigma$, which
depends on the scalar field in QMC-I while it is constant in QHD.  
However, this difference leads to 
a lot of favorable results. 

Here we consider the nucleon mass in matter further.  The nucleon mass is a 
function of the scalar field.  Because the scalar field is small 
at low density the nucleon mass may be expanded in terms of $\sigma$ as 
\bge
M_N^{\star} = M_N + \left( \frac{\partial M_N^{\star}}{\partial \sigma} 
\right)_{\sigma=0} \sigma + \frac{1}{2} \left( \frac{\partial^2 M_N^{\star}}
{\partial \sigma^2} \right)_{\sigma=0} \sigma^2 + \cdots . 
\label{nuclm}
\ene
Since the interaction Hamiltonian between the nucleon and the 
$\sigma$ field at the quark level is given by $H_{int} = - 3 g_{\sigma}^q 
\int d{\vec r} \ \overline{\psi}_q \sigma \psi_q$, the derivative of 
$M_N^{\star}$ with respect to $\sigma$ is 
$-3g_{\sigma}^q \int d{\vec r} \ \ {\overline \psi}_q \psi_q 
(\equiv -3g_{\sigma}^q S_N(\sigma))$,  
where we have defined the quark-scalar density in the nucleon 
$S_N(\sigma)$, which is itself a function of the scalar field. 
Because of a negative value of the derivative, 
the nucleon mass decreases in matter at low density.  

Furthermore, we define the scalar-density ratio $S_N(\sigma)/S_N(0)$  
to be $C_N(\sigma)$ and the $\sigma$-N coupling constant in free space    
to be $g_\sigma$ (i.e., $g_\sigma = g_\sigma(\sigma=0) 
=g_{\sigma} = 3g_{\sigma}^q S_N(0)$). 
Using these quantities, we find  
\bge
M_N^{\star} = M_N - g_{\sigma} \sigma - \frac{1}{2} g_{\sigma} 
C_N^\prime(0) \sigma^2 + \cdots . 
\label{nuclm2}
\ene
In general, $C_N$ is a decreasing function because the quark in matter is 
more relativistic than in free space.  Thus, $C_N^\prime(0)$ takes a 
negative value. If the nucleon were structureless $C_N$ would not depend on 
the scalar field.  Therefore, 
only the first two terms in the RHS of Eq.(\ref{nuclm2}) remain, 
which is exactly the same as the equation for the effective nucleon 
mass in QHD~\cite{walecka}.  

\subsection{Effect of meson structure}
It is true that not only the nucleon but also the mesons are built 
of quarks and anti-quarks, and that the mesons may change their properties 
in a nuclear medium.  
In Ref.\cite{qmc2}, we have studied the structure effects of both  
the nucleon and the mesons on the properties of finite nuclei.  
(We call this model QMC-II.)  
For further detailes on QMC-II, see Ref.\cite{qmc2} and references in 
the next section. 

\subsection{Infinite nuclear matter in QMC-I}
From the Lagrangian density Eq.(\ref{qmc-1}), we can easily find 
the total energy per nucleon  
$E_{tot}/A$ and the (constant) mean-field values of $\omega$ and 
$\rho$ (which are respectively given by baryon number 
conservation and the difference in proton and neutron 
densities).  
The scalar mean-field is given by a self-consistency 
condition, namely $(dE_{tot}/d\sigma) = 0$. 

Now we need a model for the structure of the hadrons involved to 
perform actual calculations.  We use the MIT 
bag model in static, spherical cavity approximation.  
In the present model, the bag constant $B$ and the 
parameter $z_N$ (which accounts for the sum of the c.m. and 
gluon fluctuation corrections~\cite{qmc1}) for the nucleon 
are fixed to reproduce the free nucleon mass ($M_N$ = 939 MeV) and 
the free bag radius $R_N$ = 0.8 fm.  
In the following we choose $m_q$ = 5 MeV, and set 
$m_\sigma$ = 550 MeV, $m_{\omega}$ = 783 MeV and $m_{\rho}$ = 770 MeV. 
(Variations of the quark mass and $R_N$ only lead to 
numerically small changes in the calculated results~\cite{qmc1}.)  
We then find that $B^{1/4}$ = 170.0 MeV and $z_N$ = 3.295.   

The coupling constants   
($g_{\sigma}^2$ and $g_{\omega}^2$) are fixed to fit the average 
binding energy ($-15.7$ MeV) at the saturation density ($\rho_0$) for 
symmetric nuclear matter.  Furthermore, the $\rho$-N coupling constant is 
used to reproduce the bulk symmetry energy, 35 MeV.  
We then find~\cite{qmc1}:  $g_{\sigma}^2/4\pi$ = 5.40, $g_{\omega}^2/4\pi$ 
= 5.31 and $g_\rho^2/4\pi$ = 6.93.  Note that the model gives the nuclear 
incompressibility about 280 MeV, and the variations of the nucleon bag radius 
$\delta R_N^{\star}/R_N = -0.02$, the lowest eigenvalue 
$\delta x_N^{\star}/x_N = -0.16$ and the root-mean-square radius of 
the quark wave function $\delta r_q^{\star}/r_q = +0.02$ at saturation 
density~\cite{qmc1}.  

\section{Applications}

The idea of the QMC model was first proposed by Guichon~\cite{guichon} 
in 1988, and later it has been developed by Adelaide group. 
In particular, Saito and Thomas have applied this model to various 
phenomena in nuclear physics.  We here list up those applications: 
\begin{itemize} 
\item Nuclear structure functions and the EMC effect~\cite{emc}, 
\item Equation of state (EoS) for nuclear and neutron matter~\cite{eos}, 
\item Charge symmetry breaking in nuclear matter -- 
the Nolen-Schiffer anomaly~\cite{ons}, 
\item Variation of hadron masses and matter properties in dense 
nuclear matter~\cite{qmc2,hmv}, 
\item Super-allowed Fermi beta-decay -- the unitarity problem of the 
CKM matrix element~\cite{ckm}, 
\item Properties of finite nuclei~\cite{qmc1,qmc2,finite}, 
\item {\em Naturalness}\/ in the QMC model~\cite{natural}, 
\item Hyper nuclei~\cite{hyper}, 
\item Electromagnetic form factors of the bound nucleon~\cite{elm,orbit}, 
\item In-medium Kaon properties~\cite{kaon}, 
\item Meson-nucleus bound states~\cite{mbound}. 
\end{itemize}

Furthermore, the original QMC model was improved by Jin and 
Jennings~\cite{jin}, 
in which the bag constant $B$ is allowed to decrease in nuclear matter. 
This version is called the modified QMC (MQMC) model.   The MQMC model 
has also used to calculate the properties of finite nuclei, 
and the relationship among QMC, MQMC and QHD is clarified~\cite{mqmc}. 

\section{Form factors of the bound nucleon}

As an example of recent applications, we show the in-medium 
modifications of the form factors of the nucleon~\cite{elm,orbit}.  
In QMC, the quark wave function, as well as the nucleon wave function
(both are Dirac spinors), are determined 
once a solution to equations of motion are found 
self-consistently~\cite{qmc1}.
The electromagnetic form factors for a proton bound in a specific 
orbit $\alpha$, in local density approximation,
are then simply given by
\bge
G_{E,M}^\alpha(Q^2)= \int G_{E,M} (Q^2,\rho_B(\vec{r})) 
\rho_{p\alpha}(\vec{r}) \,d\vec{r},  \label{emff} 
\ene
where $G_{E,M}(Q^2,\rho_B(\vec{r}))$ is  the density-dependent form factor 
of a ``proton'' immersed in nuclear matter with a local baryon density, 
$\rho_B(\vec{r})$ (see Ref.\cite{elm}).  
Using the calculated nucleon shell model wave functions, 
the local  baryon density and the local proton density in the specified 
orbit $\alpha$ are easily evaluated.  

The notable medium modifications of the quark wavefunction inside the bound 
nucleon include a reduction of its frequency and an enhancement 
of the lower component of the Dirac spinor.
As in earlier work~\cite{ltw}, the corrections arising from recoil and 
center of mass 
motion for the bag are made using the Peierls-Thouless
projection method, combined with Lorentz contraction of the internal
quark wave function and with the perturbative pion cloud added 
afterwards.  Additional, possible effects of off-shell form 
factors and 
meson exchange currents are ignored in the present, exploratory study. 
The resulting nucleon electromagnetic form factors agree with  experiment
quite  well in free space~\cite{ltw}, at least for momentum 
transfer less than 1 GeV$^2$.  

\begin{figure}[htb]
\begin{center}
\epsfig{file=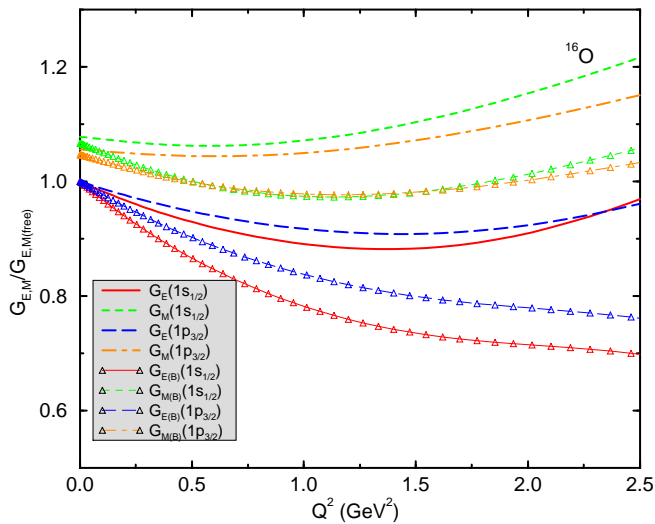,width=10cm}
\caption{Ratio of in-medium to free space electric and magnetic form factors
for the $1s$- and $1p$-orbit nucleons of $^{16}$O. 
The curves with triangles represent the corresponding ratio calculated
in MQMC with a 10\% reduction of the bag constant at $\rho_0$.}
\label{co16B.ps}
\end{center}
\end{figure}
\begin{figure}[hbt]
\begin{center}
\epsfig{file=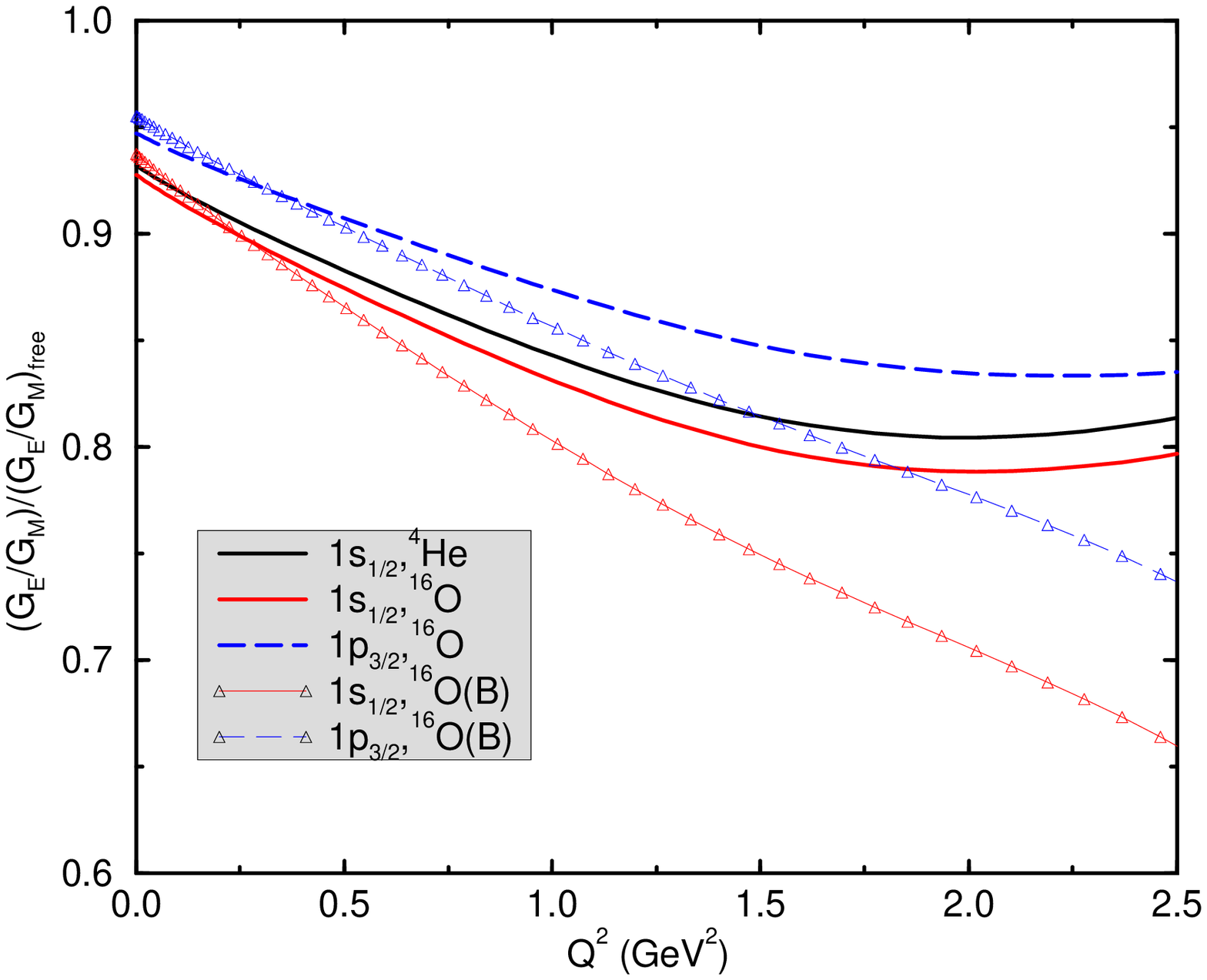,width=10cm}
\caption{Ratio of electric and magnetic form factors in-medium, divided
by the free space ratio. As in Fig.~\protect\ref{co16B.ps}, the curves 
with triangles represent the corresponding 
values calculated in MQMC.}
\label{2ratio.ps}
\end{center}
\end{figure}
In order to reduce the theoretical uncertainties, we prefer to show
the ratios
of the form factors with respect to corresponding free space values. 
In Fig.~\ref{co16B.ps} we show  
the ratios of the electric and magnetic form factors for $^{16}$O,  
which has one $s$-state ($1s_{1/2}$) and two $p$-states 
($1p_{3/2}$ and $1p_{1/2}$). 
As expected, both the electric and magnetic rms radii become slightly larger 
and the magnetic moment of the proton is also larger than the free value. 
The momentum dependence of the form factors 
for the $1s$-orbit nucleon is more supressed than those of the $1p$-states. 
This is because the inner orbit in $^{16}$O  
experiences a larger average baryon density.  
The magnetic moment for the $1s$-orbit nucleon increases by about 7\%, 
but, in the $1p$-orbits, it is reduced  by about $2 - 3$\% from the 
$1s$-orbit value.  
The difference between two $1p$-orbits is rather small.  
For comparison, we also show in Fig.~\ref{co16B.ps} 
the corresponding ratio of form factors (curves with triangles) 
using MQMC~\cite{jin,mqmc} where the bag constant is allowed to decrease 
by 10\%~\cite{elm}. 
The effect of a possible reduction in $B$ is quite large 
and severely reduces the electromagnetic form factors of the bound 
nucleon.  

From the experimental point of view, the ratio,
 $G_E/G_M$, can be derived directly from the ratio of transverse 
to longitudinal polarization of the outgoing proton, 
with minimal systematic errors. 
We find that $G_E/G_M$ runs roughly from 0.41  at $Q^2 = 0 $ to 0.28  
at $Q^2 = 1 \mbox{ GeV}^2$ for a proton in the $1s$-orbit in $^{16}$O.
The ratio of $G_E/G_M$ with respect to the corresponding free
space ratio is presented in Fig.~\ref{2ratio.ps}. 
The result for the $1s$-orbit in $^{16}$O is close to that in $^4$He  
and about $2 - 3$\% lower than that for the $1p$-orbits in $^{16}$O. 
We also find that the effect of the reduction in $B$ has a 
significant effect on this ratio of ratios, especially for larger $Q^2$. 

For completeness, we have also calculated the orbital electric and magnetic 
form factors for heavy nuclei such as $^{40}$Ca and $^{208}$Pb~\cite{orbit}.  
Because of the larger central baryon density of heavy nuclei, 
the proton electric and magnetic form factors 
in the inner orbits  
suffer much stronger medium modifications 
than those in light nuclei.
That is to say, the $Q^2$ dependence is further suppressed, while 
the magnetic moments appear to be larger.
Surprisingly, the nucleons in peripheral orbits ($1d_{5/2}$, $2s_{1/2}$, 
and $1d_{3/2}$ for $^{40}$Ca 
and $2d_{3/2}$, $1h_{11/2}$, and $3s_{1/2}$ for $^{208}$Pb) 
still endure significant medium effects, and 
comparable to those in $^4$He.  

\section{Conclusion} 

In summary, we have reviewed the quark-meson coupling model, in which 
the quark degrees of freedom are explicitly involved to describe 
not only the nuclear matter but also finite nuclei.  
Then, we have shown the electric and magnetic form factors 
for the proton bound in specific orbits for several closed-shell 
nuclei. 
Generally the electromagnetic rms radii and the magnetic moments of the 
bound proton are increased by the medium modifications.  
In view of current experimental developments, including the ability to 
precisely measure electron-nucleus scattering 
polarization observables, it should be possible 
to detect differences between the form factors in different shell model 
orbits.  The current and future experiments at TJNAF and Mainz therefore 
promise to provide vital information with which to guide and constrain 
dynamic microscopic models for finite nuclei, and perhaps 
unambiguously isolate a signature for the role of quarks.

\vskip 5mm 

I am pleased to thank Tony Thomas, Kazuo Tsushima, Ding Lu and 
Tony Williams for valuable discussions and comments.  
This work was supported by the Japan Society for the Promotion of Science. 

%
%

%
\end{document}